\documentclass[%
superscriptaddress,
twocolumn,
 amsmath,amssymb,
 aps,
prb,
]{revtex4-1}

\usepackage{graphicx}
\usepackage{dcolumn}
\usepackage{bm}

\usepackage{xcolor}
\newcommand{\notemr}[1]{{\color{black} #1}}

\begin{document}

\title{Real-time simulation of light-driven spin chains on quantum computers}

\author{Martin Rodriguez-Vega}
\affiliation{Theoretical Division, Los Alamos National Laboratory, Los Alamos, New Mexico 87545, USA }

\author{Ella Carlander}
\affiliation{Department of Physics and Astronomy, Bucknell University, Lewisburg, Pennsylvania 17837}

\author{Adrian Bahri}
\affiliation{Department of Physics, University of Michigan, Ann Arbor, MI 48109, USA}

\author{Ze-Xun Lin}
\affiliation{Department of Physics, The University of Texas at Austin, Austin, TX 78712, USA}
\affiliation{Department of Physics, Northeastern University, Boston, MA 02115, USA}

\author{Nikolai A. Sinitsyn}
\affiliation{Theoretical Division, Los Alamos National Laboratory, Los Alamos, New Mexico 87545, USA }

\author{Gregory A. Fiete}
\affiliation{Department of Physics, Northeastern University, Boston, MA 02115, USA}
\affiliation{Department of Physics, Massachusetts Institute of Technology, Cambridge, MA 02139, USA}

\date{\today}

\begin{abstract}
In this work, we study the real-time evolution of periodically driven (Floquet) systems on a quantum computer using IBM quantum devices. We consider a driven Landau–Zener model and compute the transition probability between the Floquet steady states as a function of time. We find that for this simple one-qubit model, Floquet states can develop in real-time, as indicated by the transition probability between Floquet states. Next, we model light-driven spin chains and compute the time-dependent antiferromagnetic order parameter. We consider models arising from light coupling to the underlying electrons as well as those arising from light coupling to phonons. For the two-spin chains, the quantum devices yield time evolutions that match the effective Floquet Hamiltonian evolution for both models once readout error mitigation is included. For three-spin chains, zero-noise extrapolation yields a time-dependence that follows the effective Floquet time evolution. Therefore, the current IBM quantum devices can provide information on the dynamics of small Floquet systems arising from light drives once error mitigation procedures are implemented. 
\end{abstract}

\maketitle

\section{Introduction}

The recent development of technology to carry out ultra-fast laser experiments on materials has allowed the control of topological and ordered states of matter in out-of-equilibrium settings. For example, light pulses at suitable frequencies and intensities can induce transient superconductivity~\cite{Fausti2011,mitrano2016}, anomalous Hall states in graphene \cite{McIver2020}, magnetic order switching in YIG ~\cite{Stupakiewicz2021}, and metastable ferroelectric states in SrTiO$_3$\cite{Nova1075}. Different theoretical tools have been employed to predict light-induced effects in quantum materials, mainly based on traditional theoretical and computational approaches~\cite{Oka_2019,RODRIGUEZVEGA2021168434,rudner2020_review}. 

The recent development of quantum computers and their open availability in platforms such as IBM quantum~\cite{ibmquantum} provides a new pathway to study quantum materials~\cite{Feynman1982,Lloyd1073}. Already some results have been reported~\cite{CerveraLierta2018exactisingmodel,PhysRevLett.121.170501,PhysRevLett.121.086808,Smith2019,Bassman2020,Francis2020,Fauseweh_2021}. For example, Smith, A. {\it  et al.} studied quantum quench dynamics in several spin models, and showed the presence of signatures of localization and many-body effects~\cite{Smith2019}.  Bassman {\it et al.} employed quantum devices to simulate ultra-fast control of magnetism by terahertz radiation in doped monolayer MoSe$_2$~\cite{Bassman2020}. Fauseweh and Zhu consider non-equilibrium dynamics of few spin and fermionic systems~\cite{Fauseweh_2021}, and Francis {\it et al.} implemented the computation of magnon spectra from correlation functions in spin chains~\cite{Francis2020}. Several recent articles review the state-of-the-art capabilities of near-term noisy quantum devices~\cite{Tacchino2020}, progress towards quantum simulation of quantum materials~\cite{bassman2021simulating}, and quantum algorithms~\cite{j2020quantum}. 

For the case of periodically-driven (Floquet) systems, Malz and Smith realized an effective two-dimensional Floquet lattice by driving quasi-periodically a single-qubit device and obtained topological frequency conversion~\cite{Malz_2021}. More recently, Mi {\it et al.} observed an eigenstate-ordered discrete time crystal on an array of superconducting qubits~\cite{mi2021observation}. These works provide compelling evidence that the current quantum devices already yield information regarding dynamical aspects of quantum materials. 

We consider two classes of periodically-driven models in this work: a driven Landau-Zener model and light-driven spin chains. The former model serves as a one-qubit system example. The later models arise as effective representations of laser irradiated Hubbard models at half-filling and are directly relevant to describing quantum materials. Furthermore, we examine both the cases of light coupling with the electrons and phonon degrees of freedom. Thus, in contrast to previous works approaching Floquet systems, we discuss the solution of periodically-driven models with direct applications to quantum materials out of equilibrium. 

We implement the driven Landau-Zener model in a single-qubit IBM device and find that upon applying a simple error correction procedure, the Floquet states are well reproduced. For the light-driven spin chain, we consider systems with two and three spins and obtained effective Floquet time evolution from the quantum devices upon the implementation of zero-noise extrapolation. We notice that the error correction procedures considered in this work are not quantum error correction in conventional sense, but rather special tools that correct for systematic errors in the quantum processor and thus make more precise quantum gates. Therefore, our work shows that current quantum devices can realize Floquet states in systems with few spins. 

The rest of the paper is organized as follows. In Sec. II, we provide a brief review of Floquet theory. In Sec. III, we study the driven Landau-Zener model implemented in a single-qubit device. In Sec. IV, we look at light-driven spin chains, assuming that light couple to the electrons. In Sec. V, we consider spin chains with time dependence arising from driven phonons. Finally, in Sec. VI we present our conclusions and outlook. 

\section{Review of Floquet theory}

First, we briefly review the general aspects of Floquet theory. For a complete review with applications to quantum materials, see Refs.\cite{Oka_2019,RODRIGUEZVEGA2021168434,rudner2020_review}. Experimentally, Floquet states have been observed in the topological insulator Bi$_2$Se$_3$~\cite{wang2013}, Fermi gases~\cite{PhysRevA.98.013615}, photonic platforms~\cite{rechtsman2013}, and ultra-cold atoms~\cite{Wintersperger_2020}.

The starting point of study in Floquet systems is a time-dependent Hamiltonian satisfying the periodicity condition $\mathcal H(t+2\pi/\Omega) = \mathcal H(t)$, where $\Omega$ is the drive frequency. For example, in quantum materials such a time dependence can originate from laser excitation (the electrical field of the light oscillates with the frequency of the light and therefore causes the material Hamiltonian to oscillate with the same frequency). 

The Floquet theorem \cite{ASENS_1883_2_12__47_0} indicates that the wavefunctions of such time-periodic Hamiltonians can be written as $| \psi (t) \rangle = e^{i \epsilon t} |\phi(t) \rangle$, where $|\phi(t + 2\pi/\Omega) \rangle = |\phi(t) \rangle$ share the periodicity with the Hamiltonian and $\epsilon$ is the quasienergy, defined modulo integer multiples of $\hbar \Omega$. The Floquet-Schr\"odinger equation takes the form
\begin{equation}
[\mathcal H(t)-i\partial_t] |\phi(t)\rangle = \epsilon |\phi ( t)\rangle,
\end{equation}
which can be solved in the time-domain by diagonalizing the Floquet time-evolution operator $U(T)$, with 
\begin{equation}
 U(t) = \mathcal T \exp \{-i \int^t \mathcal H(s) ds \},   
\end{equation}
and $\mathcal T \exp$ is a time-ordered exponential. The full dynamics of the wavefunction can be obtained as $|\psi(t) \rangle = U(t) |\psi(0) \rangle$. Alternatively, the Floquet-Schr\"odinger equation can be solved in the extended-space by exploiting the periodicity of the Floquet states~\cite{Eckardt_2015, PhysRevA.7.2203}. This frequency-domain picture is suitable for analytical approximation schemes and numerical implementation in classical devices. However, the time-domain approach is naturally suited for implementation in quantum devices. 

In the following sections, we implement time-dependent periodic Hamiltonians in state-of-the-art IBM quantum devices and study the development of Floquet states. 

\section{Driven Landau–Zener model}

In this section, we consider the case of single spin-1/2 model. The authors of Ref. \cite{Fauseweh_2021} considered a model for a spin-1/2 in a time-dependent magnetic field. In this work, we consider the Hamiltonian for a model that we will call driven Landau-Zener model, given by
\begin{equation}
H(\tau)=a f(\tau) \sigma_{x}+b \sigma_{y},
\end{equation}
where $\tau = \Omega t$, $\Omega$ is the drive frequency, $a,b > 0$, $\sigma_i$ are the Pauli matrices, and $f(\tau)$ is a time-dependent function with the property $f(0) = - f(\pi)$. We consider $f(\tau) = \cos(\tau)$. In the case $b=0$, the system admits an exact solution \notemr{for the time evolution operator which is given by $U(\tau) = e^{-i a \sin(\tau) \sigma_x/\Omega }$. However, this is not the case for arbitrary $b$ (which is the case we consider in this work) and we need to introduce approximations for the time evolution operator $U(\tau)$ as discussed below.} For an extensive discussion on the driven Landau–Zener model in the context of low-frequency Floquet perturbation theory, see Ref. \cite{Rodriguez_Vega_2018} and the references therein. 

We consider the parameters $b/\Omega=1$, and $a/b=25$, and study the transition probability between the energy levels as a function of time 
\begin{equation}
  P_{\pm}(\tau)=\left|\left\langle\psi_{\mp}(\tau)| U(\tau)| \psi_{\pm}(0)\right\rangle\right|^{2},  
\end{equation}
 where $|\psi(\tau)_{\pm} \rangle = U(\tau) |\psi(0)_{\pm} \rangle $, $U(\tau)$ is the time-evolution operator and $|\psi(0)_{\pm} \rangle$ is an arbitrary initial state. 

To evaluate the transition probability $P_{\pm}(\tau)$ in a quantum device, we initialize the qubit in the state $|\psi(0)_{\pm} \rangle = 1/\sqrt{2}(1,\pm1)$ by applying a single-qubit rotation gate about the $y$-axis, $R_y(\theta)$, to the default qubit state. \notemr{A general propagator $U(\tau)$ can be constructed as follows. First, we define a grid in time domain with step size $\Delta t$, during which the Hamiltonian $H(\Delta t)$ is assumed to be constant. Then, the time-evolution operator can be approximated as}
\begin{equation}
\notemr{U(N \Delta t)=\prod_{n=0}^{N-1} \prod_{X} e^{-i H_{X}(n \Delta t) \Delta t}+\mathcal{O}(\Delta t),}
\end{equation}
\notemr{where the set $\{X\}$ is given by non-commuting Hamiltonian terms which can be expressed in terms of gates~\cite{Smith2019,Bassman2020,Fauseweh_2021}. $N$ is the number of time steps considered. For the driven Landau-Zener, we can compute the Hamiltonian exponential analytically and write the approximate time-evolution operator as $U(N \Delta t)=\prod_{n=0}^{N-1} e^{-i H(n \Delta t) \Delta t}$, with} 
\begin{equation}
\notemr{e^{-i \Delta t H(s)}= \begin{pmatrix}
\cos(\eta(s) \Delta t ) & -\sin(\eta(s) \Delta t ) \nu(s)\\
\sin(\eta(s) \Delta t )\nu^*(s) & \cos(\eta(s) \Delta t ) 
\end{pmatrix},}
\end{equation}
\notemr{$\eta(s) = \sqrt{b^2+a^2\cos^2(s)}$, and $\nu(s)=\left(b+i a \cos(s) \right)/\eta$ with $|\nu(s)|=1$. Thus, the matrix structure of each time step $e^{-i \Delta t H(s)}$ corresponds to a universal $U3(\theta, \phi, \lambda)$ gate with time-dependent Euler angles $\theta(s)/2 = \eta(s) \Delta t$, $\phi(s) = -a/b \cos(s)$, and  $\lambda(s) = a/b \cos(s).$ The circuit for a given time evolution time interval is then given by a set of $U3(\theta, \phi, \lambda)$ gates. The standard Qiskit function ``transpile" can optimize the circuit and write it as a single $U3$ gate.}

\begin{figure}[t]
	\centering
		\includegraphics[width=8.5cm]{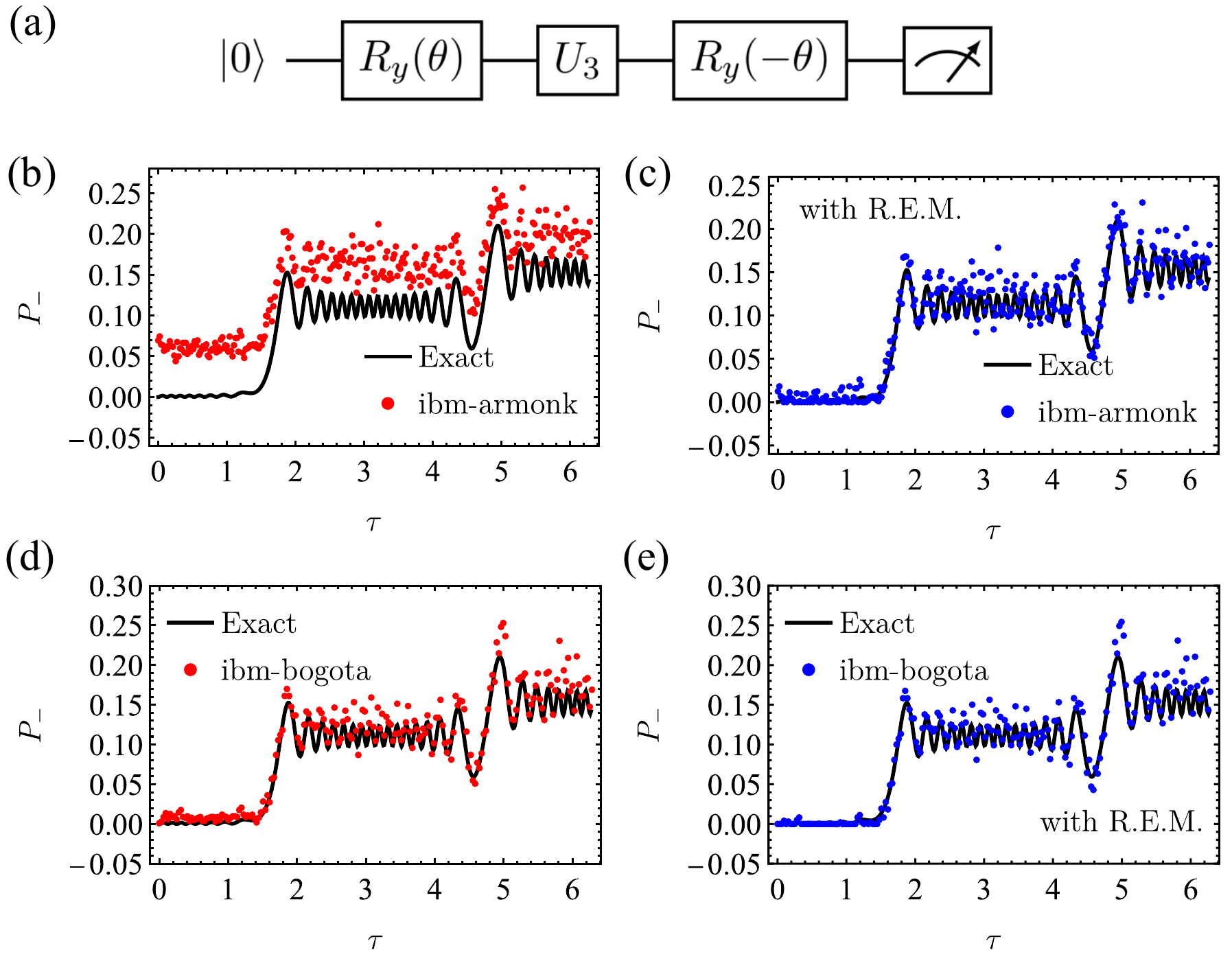}
		\caption{(Color online) (a) Quantum circuit used to solve the driven Landau-Zener model, created with the Quantikz package~\cite{kay_2018}. $U_3$ represents a general one-qubit rotation gate with three Euler angles obtained with Qiskit. $R_y(\theta)$ represents a single-qubit rotation about the $y$-axis (b) Transition probability obtained from the single-qubit ibmq-armonk device (red dots) compared with the exact solution (black solid line). \notemr{$\tau = t \Omega$ is the re-scaled time, plotted over one period}. There is an overall shift between the exact result and the quantum computer result. Panel (c) shows the same data as in (b), but including readout error mitigation (R.E.M.), implemented as described in the text. This simple error mitigation procedure improves the quality of the solutions obtained from the ibmq-armonk device.
		In panels (d) and (e) we show analogous calculations to panels (b)-(c), but obtained with the ibm-bogota device. \notemr{For sampling, in all our experiments we consider $1024$ repetitions of each circuit.}
		}
\label{fig:fig1}
\end{figure}

Finally, we apply a rotation $R_y(-\theta)$ to measure the probability to find the qubit in the states $|\psi(0)_{\pm} \rangle$. From these measurements, we construct the transition probability $P_{\pm}(\tau)$. We show the quantum circuit in Fig.~\ref{fig:fig1}(a). 

In Fig.~\ref{fig:fig1}(b), we show the exact solution obtained numerically compared with the results from the single-qubit IBM quantum computer ibmq-armonk. For sampling, in all our experiments we consider $1024$ repetitions of each circuit. At the time this device was accessed, the average readout error was $2.689\times 10^{-2}$, $T_1 = 176.97$~$\mu$s (decay time from the excited state down to the ground state), and $T_2 = 229.80$~$\mu$s (coherence time). The results from ibmq-armonk follow the trends of the exact solution for the transition probability $P_{\pm}(\tau)$, apart from an overall shift. 
One source of error in the IBM quantum computer stems from the final readout procedure~\cite{qiskitreadoutwerror}, wherein some states `1' are read as `0' and vice-versa for the single qubit case. If we arrange the possible outputs in the basis $\{1, 0\}$, we can express the results as a vector $C_{\text{noisy}}$. For $\mathcal{N}$ qubits, this vector has length $2^\mathcal{N}$. The ideal result can be written as $C_{\text{noisy}}=M C_{\text{ideal}}$, where $M$ is a calibration matrix that is obtained by performing measurements of all possible outcomes. In a perfect device, $M$ is an identity matrix. For ibmq-armonk, we find the calibration matrix
\begin{equation}
M = \left[\begin{array}{ll}
0.94141 & 0.08105 \\
0.05859 & 0.91895
\end{array}\right].
\end{equation}

By applying this transformation to the noisy outcome, we obtain the results in Fig.~\ref{fig:fig1}(c), which are in good agreement with the exact solution. We obtain similar results in other IBM devices, such as the five-qubit ibmq-bogota device, as shown in Fig.~\ref{fig:fig1}(d)-(e). Panel (d) shows the results obtained directly from the ibmq-bogota, while panel (e) shows the results including readout error mitigation, as described above. At the time this five-qubit device was accessed, the average readout error was $2.090\times 10^{-2}$, $T_1 = 68.65$~$\mu$s, and $T_2 = 96$~$\mu$s. These results indicate that Floquet states are accessible for single-spin Hamiltonians, in both quantum devices considered. 

\section{Light-driven spin chains}

\begin{figure}[t]
	\centering
		\includegraphics[width=8.5cm]{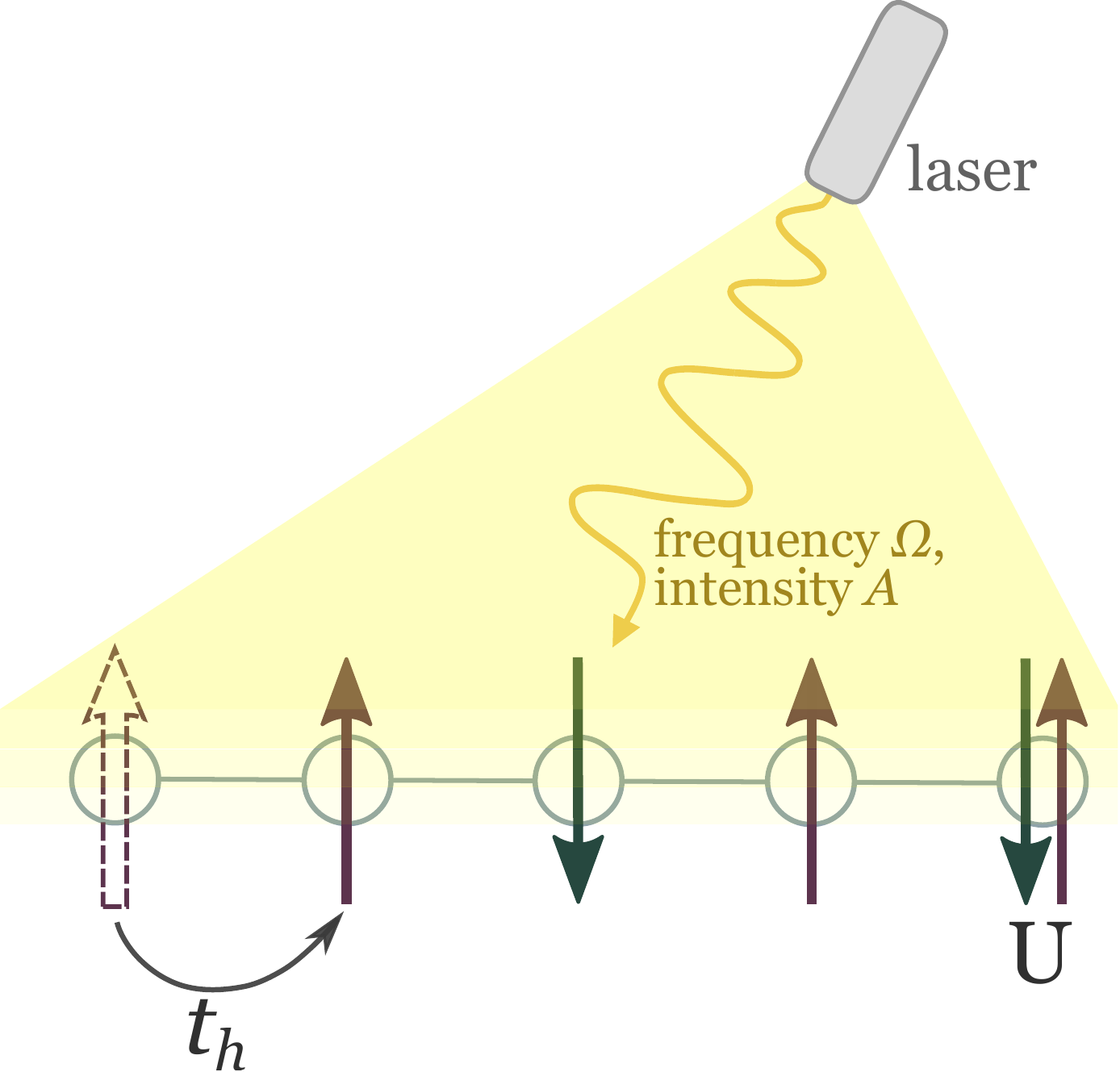}
		\caption{(Color online) Sketch of a periodically driven Hubbard model at half filling. Here $t_h$ is the hopping amplitude, and $U$ the Coulomb interaction. The laser is described by its frequency $\Omega$ and intensity $A$.}
\label{fig:cartoon}
\end{figure}

In this section, we consider a light-driven   Hubbard model at half-filling defined on a one-dimensional chain. A sketch of the system is shown in Fig.~\ref{fig:cartoon}. The Hamiltonian is given by~\cite{Hejazi_2019,Mentink2015,Claassen2017,bukov2016}
\begin{equation}
H(t)=-\sum_{i \sigma}\left(t_{h} e^{i A \sin \Omega t} c_{i \sigma}^{\dagger} c_{i+1 \sigma}+\text { H.c. }\right)+U \sum_{i} \hat{n}_{i \uparrow} \hat{n}_{i \downarrow},
\label{eq:mottlight}
\end{equation}
where $t_{h}$ is the hopping amplitude between nearest-neighbor sites, \notemr{$c^\dagger_{i\sigma}$ creates an electron at lattice site $i$ with spin $\sigma$,} $U$ is the on-site Coulomb interaction, and $\Omega$ is the frequency of the light. We consider the limit $U \gg t_{h}$. Here, $A=e E_{0}a_0 / (\hbar \Omega)$, where ${E}_{0}$ is the peak laser electric field, $\Omega$ is the laser frequency, and $a_0$ is the nearest-neighbor distance. 

The authors of Refs.~\cite{Mentink2015,Claassen2017,bukov2016} derived an effective Floquet (time-independent) spin Hamiltonian from $H(t)$,  Eq.\eqref{eq:mottlight}, via Brillouin-Wigner perturbation theory or a Schrieffer-Wolff transformation. On the other hand, the authors of Ref.\cite{Hejazi_2019} used time-dependent
second-order perturbation theory, and derived an effective model in the time-domain. We follow their prescription and find the effective model
\begin{equation}
H_s(t) = \sum_{\langle i j\rangle} J(t)\mathbf{S}_{i} \cdot \mathbf{S}_{j},
\label{eq:hams}
\end{equation}
with \notemr{lattice-site independent, time-dependent exchange interaction}
\begin{align}\nonumber
J(t)&=\Re \Biggl[\sum_{\alpha,\beta=-\infty}^{\infty} e^{i(\alpha-\beta) \Omega t} [\mathcal{J}_{\alpha}\left(A \right) \mathcal{J}_{-\beta}\left(-A\right)\\
& +
\mathcal{J}_{\alpha}\left(-A \right) \mathcal{J}_{-\beta}\left(A\right)
]\left(\frac{2 t_{h}^{2}}{U-\beta \Omega}\right) \Biggr]
\end{align}
valid when the condition $|U-\beta \Omega| \gg t_h$ is satisfied, \notemr{where $\alpha,\beta$ label the Fourier modes}. \notemr{$\mathcal{J}_{\mathcal{\alpha}}\left(A \right)$ corresponds to the $\alpha$-th Bessel function of the first kind}. Taking a time average, one finds the effective Floquet exchange interaction $J_{F} = \sum_{\beta} \mathcal{J}_{\beta}\left(A \right)^2 4t^2_h/(U-\beta\Omega)$, valid for laser frequencies larger than the exchange energy and extensively discussed in the literature~\cite{Mentink2015,Claassen2017,bukov2016,Hejazi_2019}. 

\begin{figure}[t]
	\centering
		\includegraphics[width=8.5cm]{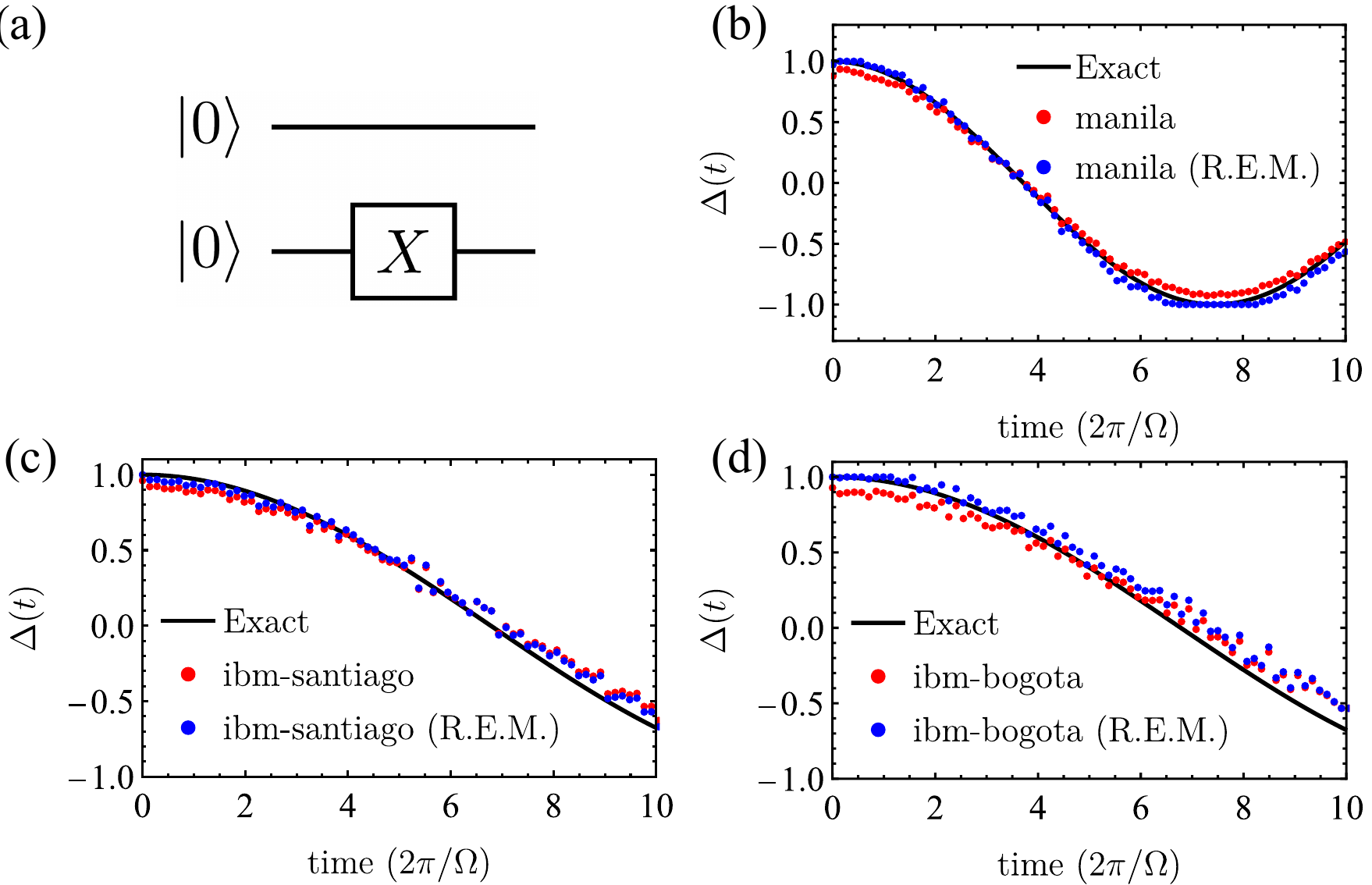}
		\caption{(Color online) (a) Quantum circuit defining the AFM initial state, created with the Quantikz package~\cite{kay_2018}. The X-gate corresponds to $X =|0\rangle\langle 1|+| 1\rangle\langle 0|$. (b) AFM order parameter $\Delta(t)$, Eq.\eqref{eq:Delta}, as a function of time for a two-spin chain with a time-evolution governed by $J=4t_h^2/U$ for $U=10$. The results obtained in the ibm-manila quantum device are shown in red, with error mitigated results in \notemr{blue}. \notemr{To set the time scale, we use the frequency $\Omega = 6.0$.} (c) ((d)) $\Delta(t)$ for a light-driven spin chain with $\Omega = 6.0$ and $A=2.8$ obtained in the quantum device ibm-santiago \notemr{(ibm-bogota)}. The shown exact solution corresponds to the effective time evolution governed by $J_F$, which the quantum device captures for the ten periods considered. 
		}
\label{fig:fig2}
\end{figure}

Next, we implement $H_s(t)$ in quantum devices. For comparison, we also consider the case without light, described by the time-independent Hamiltonian $ 
H_s = \sum_{\langle i j\rangle} J \mathbf{S}_{i} \cdot \mathbf{S}_{j}$, with $J = 4 t^2_h/U$. The procedure follows the same steps as for the single-spin case. First, we define an initial state given by the groundstate antiferromagnetic configuration. We show the case for two qubits in Fig.~\ref{fig:fig2}(a). Then, we apply gates to the qubits to simulate the time-evolution operator as described below, and finally we perform a measurement. 

For the spin chain case, we are required to introduce approximations to write the time-evolution operator $U(t)$ as a sequence of two-qubit gates, including CNOT gates. CNOT gates, typically have larger errors compared with single-qubit gates in current quantum devices. The theory for the quantum simulation of time-dependent Hamiltonians is introduced in Ref. \cite{PhysRevLett.106.170501}, and Refs. \cite{Smith2019,Bassman2020,Fauseweh_2021} discussed implementations in detail.\notemr{ The procedure to compute the time-evolution operator was outlined in Section III Driven Landau-Zener model. For completeness, we repeat here the discussion.} First, we define a grid in time domain with step size $\Delta t$, during which the Hamiltonian $H_s(\Delta t)$ is assumed to be constant. Then, the time-evolution operator can be approximated as
\begin{equation}
U(N \Delta t)=\prod_{n=0}^{N-1} \prod_{X} e^{-i H_{X}(n \Delta t) \Delta t}+\mathcal{O}(\Delta t),
\end{equation}
where the set $\{X\}$ is given by non-commuting Hamiltonian terms which can be expressed in terms of gates~\cite{Smith2019,Bassman2020,Fauseweh_2021}. 

We start with a minimal chain with $\mathcal{N}=2$ spins . First, we consider the time evolution of the antiferromagnetic groundstate $| \psi(0) \rangle = |\uparrow \downarrow \rangle$ under the time-independent Hamiltonian $ 
H_s = \sum_{\langle i j\rangle} J \mathbf{S}_{i} \cdot \mathbf{S}_{j}$, with $J = 4 t^2_h/U$. We use the parameter $U=10$ in units of the hopping amplitude $t_h$. In Fig.~\ref{fig:fig2}(b), we plot the result from the IBM quantum device ibm-manila for the time-dependent antiferromagnetic order parameter
\begin{equation}
\Delta(t) \equiv 1 / \mathcal{N} \sum_{i} (-1)^i \sigma_{i}^{z}(t),
\label{eq:Delta}
\end{equation}
where $\mathcal{N}$ is the number of spins in the chain. We select a noise-adaptive layout to associate the physical qubits with the circuit virtual qubits~\cite{murali2019noiseadaptive}. For comparison, we display the exact solution obtained numerically. As in the case for the single-qubit, implementing readout error mitigation improves the solution and we obtain a good description of the dynamics for the times considered. In this case, calibration measurements of the states $00,10,01,11$ are needed to implement the readout error mitigation protocol. We show the calibration data for the day the device was accessed in the Appendix B. For each point in time, we collect data from $1024$ experiments.

Next, we review the time evolution under the effect of light, as described by Eq. \eqref{eq:hams}. We use the drive parameters $\Omega = 6$ (in units of the hopping amplitude $t_h$) and $A=2.8$. In Fig.~\ref{fig:fig2}(c) ((d)), we plot the time-dependent antiferromagnetic order parameter as a function of time, for ten drive cycles. The solid curve corresponds to the effective Floquet time evolution, governed by the effective Floquet exchange interaction $J_F$, closely matching the solution from the quantum simulation performed in the IBM devices ibm-santiago (ibm-bogota). Therefore, for the two spin chains, the effective Floquet dynamics is obtained in the quantum device. \notemr{The user-designed circuit is shown in Appendix A for a given time step, along with the circuit obtained after Qiskit optimization}. We note that the standard Qiskit circuit optimization routines lead to a constant-depth circuit with three CNOT gates for all the times considered for the two-spin case. 

Now we consider chains with $\mathcal{N}=3$ spins. As the number of spins in the chain increases, more CNOT gates are required to simulate the dynamics of the light-driven spin chain. Since CNOT gates present more significant errors compared with single-qubit gates, accurate quantum simulations become more challenging. We employ a symmetric Trotter decomposition~\cite{Smith2019} (\notemr{the circuit for the first time step is shown in Appendix C}) with noise-adaptive layout mapping from virtual to physical qubits~\cite{murali2019noiseadaptive}. \notemr{We set the number of symmetric Trotter steps to $N=8$, enough for convergence, in the interval considered for a simulator without simulated noise. We show the results for $N=4, 8$ steps in Appendix \ref{App:3spinsConvg}.} Besides readout error mitigation, we consider a zero-noise extrapolation scheme, as implemented in the Mitiq package~\cite{larose2021mitiq}. We use random gate folding and a linear extrapolation method with two noise scaling factors. For each point in time, we collect data from $6144$ experiments. 

\begin{figure}[t]
	\centering
		\includegraphics[width=8.5cm]{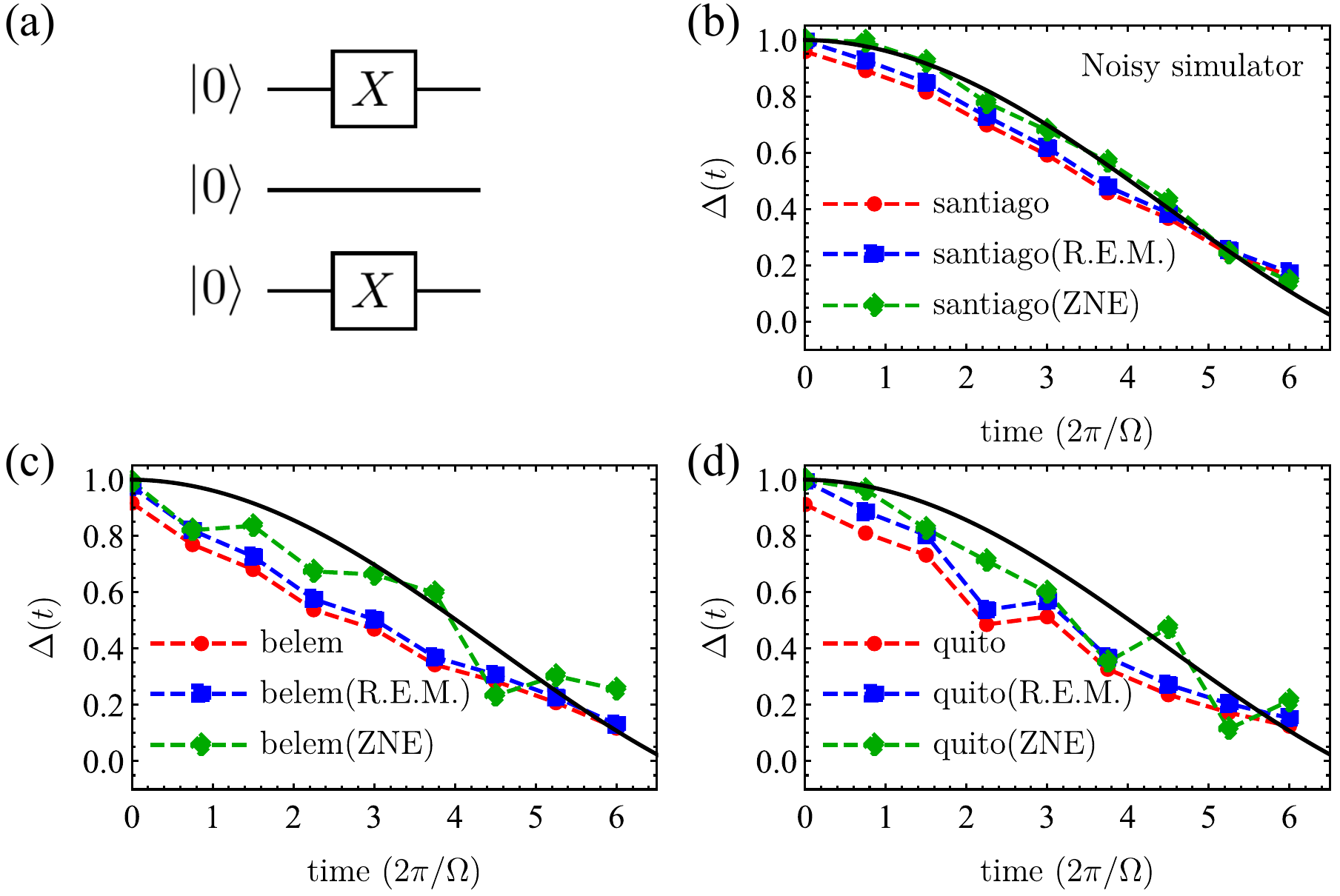}
		\caption{(Color online) (a) Quantum circuit defining the three-site AFM initial state, created with the Quantikz package~\cite{kay_2018}. (b) AFM order parameter $\Delta(t)$, Eq.\eqref{eq:Delta}, as a function of time for a light-driven three-spin chain with $U=10$, $A=2.8$, and $\Omega=6$. The results were obtained in a noisy simulator. The black line is the effective Floquet exact solution, the red circles are the direct results from the noisy simulator. The blue squares include readout error mitigation (R.E.M), and the green diamonds incorporate zero-noise extrapolation (ZNE). Panels (c) and (d) show the corresponding results obtained in the actual quantum devices ibm-belem and ibm-quito. }
\label{fig:fig3}
\end{figure}

Fig.~\ref{fig:fig3}(a) shows the AFM initial qubit state, which we then propagate in time by applying the time-evolution operator. First, we use a simulator including an approximation to the errors of the actual quantum device ibm-santiago. The average device properties of the simulator are: $T_1=124.04$~$\mu$s, $T_2=107.11$~$\mu$s, CNOT error $6.02\times 10^{-3}$, and averaged readout error $0.014$. We show the results in Fig. \ref{fig:fig3}(b). The solid line corresponds to the exact results for the effective Floquet exchange interaction obtained numerically via exact diagonalization. The red dots are the results obtained directly from the noisy simulator. The blue squares take readout error mitigation into account, and the green diamonds include zero noise extrapolation combined with readout error mitigation. The results including both error correction procedures follow the Floquet exact solution well. Next, we show results from actual quantum devices.  In Fig.~\ref{fig:fig3}(c) and (d), we summarize our results from two experiments conducted in the IBM quantum devices ibm-quito and ibm-bogota. The results, including zero noise extrapolation, show an improvement compared with the results obtained directly from the quantum devices \notemr{and follow the trend well in the time interval considered.} 

For larger spin chains with $\mathcal{N}=4$ and $\mathcal{N}=5$ spins, we found that we can obtain good approximations to the exact solution in some noisy simulators upon implementation of zero-noise extrapolation. However, running the quantum circuits in the actual quantum devices yields results that no longer follow the effective Floquet solution \notemr{consistently across quantum devices} with the methods considered in this work. \notemr{ We discuss in detail the $\mathcal{N}=4$-spin chain results in Appendix E, and the $\mathcal{N}=5$-spin chain results in Appendix F. Current research efforts in the community look for more efficient quantum algorithms. For example, Ref. \cite{bassman2021constantdepth} shows that there are some time-dependent Heisenberg Hamiltonians with $\mathcal{N}$ spins (not including the light-driven models here discuseed) that admit constant-depth circuits. This, in principle, would allow long-time simulations for longer spin chains.}

In the next section we investigate time-dependent spin Hamiltonians arising from a time-dependent bond distance, modeled as a time-dependent hopping amplitude. This class of time-dependence could occur, for example, from driven phonons in the harmonic regime.

\section{Phonon-driven spin chain}

\begin{figure}[t]
	\centering
		\includegraphics[width=7.0cm]{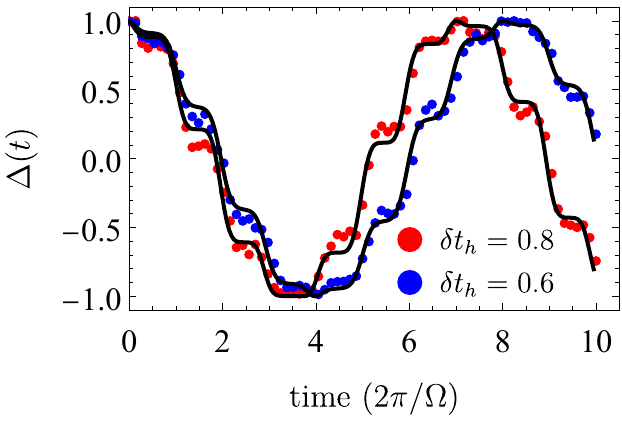}
		\caption{(Color online) Antiferromagnetic order parameter $\Delta(t)$, Eq.\eqref{eq:Delta}, as a function of time for a phonon-driven two-spin chain for $\Omega=4.0$ and two values for $\delta t_h$. The black line corresponds to the exact solution, while the dots correspond to quantum simulations performed in the ibm-lima device, including readout error mitigation (R.E.M.).}
\label{fig:fig5}
\end{figure}

In this section, we consider a Hubbard model at half-filling with time-dependent hopping amplitude, 
\begin{align}\nonumber
H(t)&= -\sum_{i \sigma}[t_h+\delta t_h \cos (\Omega t+\phi)]\left(c_{i \sigma}^{\dagger} c_{i+1 \sigma}+ H.c.\right)\\
&+U \sum_{i} \hat{n}_{i \uparrow} \hat{n}_{i \downarrow},
\label{eq:mottphonon}
\end{align}
where $\delta t_h$ corresponds to the change in the hopping amplitude arising from variation in the bond length due to a driven harmonic phonon with frequency $\Omega$. Employing time-dependent second-order perturbation  theory, we arrive at the time-dependent effective spin model
\begin{equation}
H_s(t) = \sum_{\langle i j\rangle} J(t)\mathbf{S}_{i} \cdot \mathbf{S}_{j},
\end{equation}
with time-dependent exchange interaction
\begin{align}\nonumber
J(t)&=4\left[t_h+\delta t_h \cos( \Omega t)\right]\times \\&
\left[\frac{t_h}{U}+\frac{\delta t_h}{2} \frac{e^{i \Omega t}}{(U+\Omega)}+\frac{\delta t_h}{2} \frac{e^{-i \Omega t}}{(U-\Omega)}\right],
\end{align}
valid also for $|U - \Omega | \gg t_h$. In this case, when the frequency is larger than the static exchange interaction $J=4 t_h^2/U$, the effective Floquet exchange interactions is given by $J_{F}=\frac{4 t^{2}}{U}+(\delta t)^{2} \frac{1}{U-\omega}+(\delta t)^{2} \frac{1}{U+\omega}$. 

For this model, we consider the frequency $\Omega = 4 $, and consider and two representative values for $\delta t_h$. We show our results in Fig.~\ref{fig:fig5}. The black lines correspond to the full exact time-dependent solution obtained by exact diagonalization. The blue and red dots are the results obtained from the quantum device ibm-lima, including readout error mitigation. The calibration data is shown in Appendix G. As for light-driven spin chains, the phonon-driven two-spin chain is well modeled in quantum devices. 

\section{Conclusions}

This work studied the implementation of periodically-driven Hamiltonians in IBM quantum devices, currently accessible to the public. We considered a driven Landau-Zener model and showed that the Floquet states are obtained, as shown by the transition probability as a function time within one period upon implementing readout error mitigation. We also studied time-dependent Hamiltonians describing light- and phonon-driven spin chains with two, three, four and \notemr{five} spins. We found that accurate results can be obtained for ten drive cycles using readout error mitigation for two spin chains. \notemr{For three spin chains, we implemented zero-noise extrapolation to improve the performance of the quantum devices, and similarly for four and five spin chains in noisy quantum simulators. The results for four- and five-spin chains from quantum devices are less accurate.} Therefore, current quantum devices can describe the dynamics of small spin chain models driven by light and phonons. For future work, it would be interesting to consider the effect of spin-orbit coupling in spin models.

\section{Acknowledgements}
We thank Adam Smith and Lindsay Bassman for helpful discussions. This research was primarily supported by the National Science Foundation through the Center for Dynamics and Control of Materials: an NSF MRSEC under
Cooperative Agreement No. DMR-1720595, with additional support from NSF DMR-1949701 and NSF DMR-2114825. M. R-V. and N.A.S. were supported by LANL LDRD Program and by the U.S. Department of Energy, Office of Science, Basic Energy Sciences, Materials Sciences and Engineering Division, Condensed Matter Theory Program.

\section*{References}

%

\clearpage

\onecolumngrid
\appendix

\section{Example circuit for two-spin chains.}

\begin{figure}[ht]
	\centering
		\includegraphics[width=15.0cm]{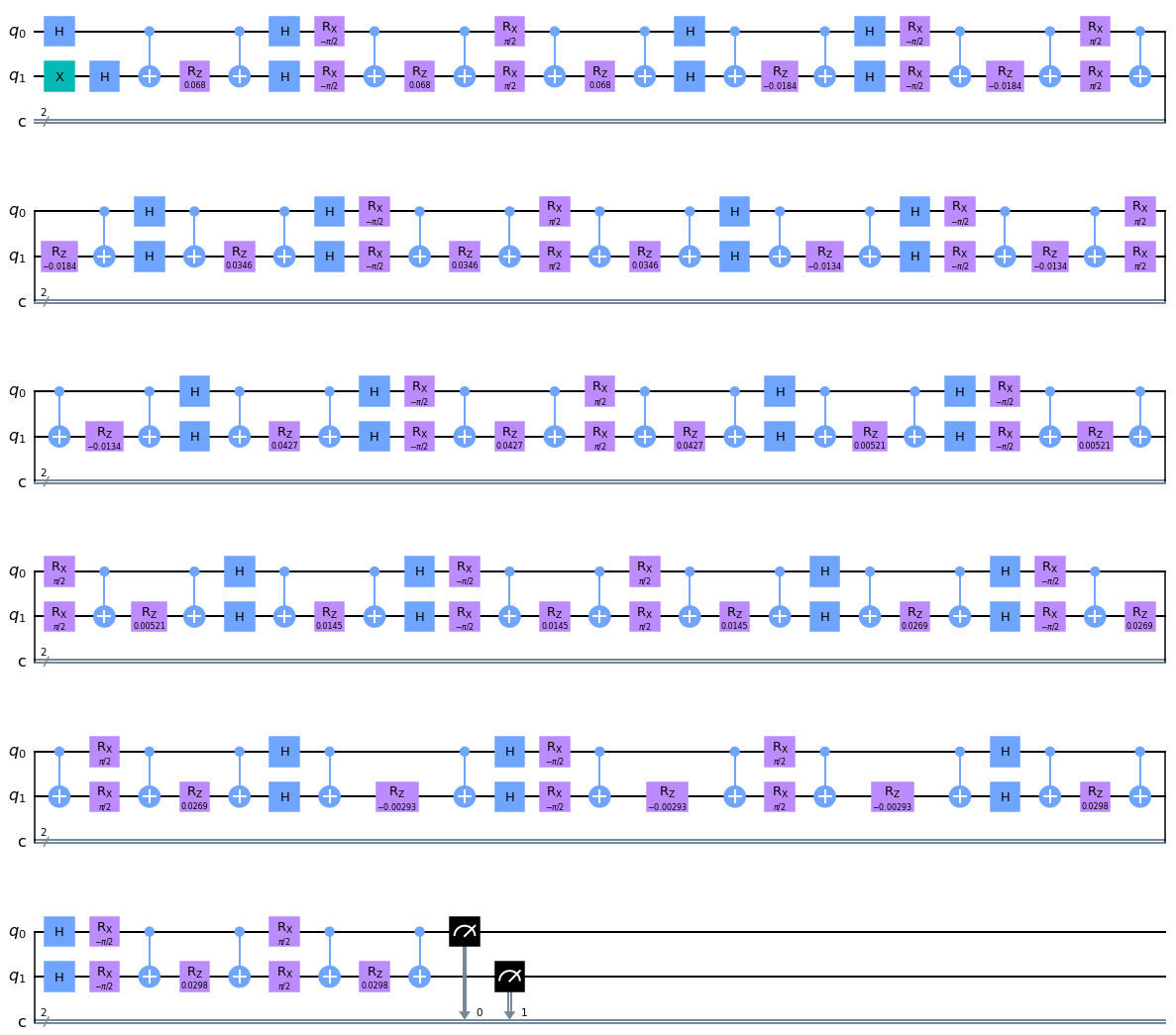}
		\caption{(Color online) User-designed quantum circuit used to simulate a light-driven two-spin chain for time step $n=10$. No Qiskit optimization was used. The drive parameters are the same as in Fig. 3.}
\label{fig:appA1}
\end{figure}

\begin{figure}[ht]
	\centering
		\includegraphics[width=8.5cm]{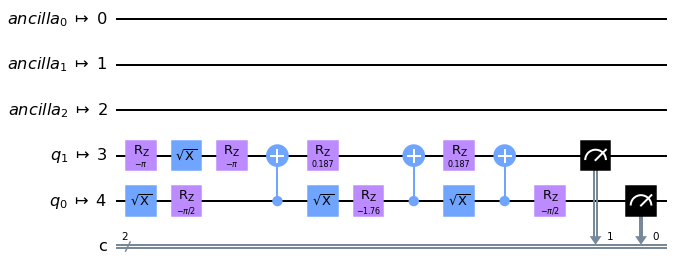}
		\caption{(Color online) Quantum circuit used to simulate a light-driven two-spin chain for time step $n=10$, obtained from the standard Qiskit optimization routines, and the noise-adaptive layout method. The user-designed circuit is shown in Fig. \ref{fig:appA1}. The drive parameters are the same as in Fig. 3.}
\label{fig:appA2}
\end{figure}

\begin{figure}[ht]
	\centering
		\includegraphics[width=8.5cm]{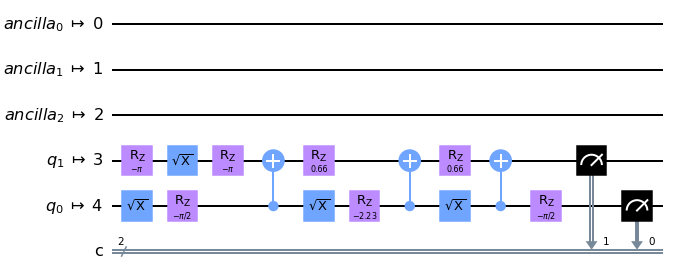}
		\caption{(Color online) Quantum circuit to simulate a light-driven two-spin chain for time step $n=40$, obtained from the standard Qiskit optimization routines, and the noise-adaptive layout method. The drive parameters are the same as in Fig. 3.}
\label{fig:appA3}
\end{figure}

In this appendix, we show the user-designed quantum circuit used to simulate a light-driven spin chain, for the time step $n=10$ (Fig. \ref{fig:appA1}), compared with the circuit optimized using the standard Qiskit optimization routines (Fig. \ref{fig:appA2}). The number of CNOT gates obtained for this model after optimization is reduced to only three, independent of the time step. As an example, in Fig. \ref{fig:appA3}, we show time step $n=40$. This simplification is only obtained via the Qiskit optimization for two-spin chain case. However, Ref. \cite{bassman2021constantdepth} shows that constant-depth circuits can be obtained for some time-dependent Heisenberg Hamiltonians. The light-driven spin chain models here considered do not belong to such classes, but further research could lead to such extensions. 

\clearpage
\section{Quantum device calibration data for two-spin chains}

\begin{table}[ht]
\begin{tabular}{|c|c|c|c|c|c|c|c|c|c|}
\hline
Qubit & T1 (us) & T2 (us) & Freq. (GHz) & Anharm. (GHz) & Readout error & ID error & $\sqrt{x}$ (sx) error & Pauli-X error & CNOT error                   \\ \hline
Q0    & 91.79   & 93.83   & 4.963       & -0.34335      & 3.080e-2      & 2.260e-4 & 2.260e-4      & 2.260e-4      & 0\_1:1.442e-2                \\ \hline
Q1    & 101.22  & 43.76   & 4.838       & -0.34621      & 1.205e-1      & 5.350e-4 & 5.350e-4      & 5.350e-4      & 1\_2:1.483e-2; 1\_0:1.442e-2 \\ \hline
Q2    & 162.27  & 23.27   & 5.037       & -0.34366      & 3.710e-2      & 3.389e-4 & 3.389e-4      & 3.389e-4      & 2\_3:7.611e-3; 2\_1:1.483e-2 \\ \hline
Q3    & 136.86  & 56.73   & 4.951       & -0.34355      & 1.740e-2      & 1.804e-4 & 1.804e-4      & 1.804e-4      & 3\_4:7.120e-3; 3\_2:7.611e-3 \\ \hline
Q4    & 123.49  & 36.07   & 5.066       & -0.34211      & 4.250e-2      & 6.500e-4 & 6.500e-4      & 6.500e-4      & 4\_3:7.120e-3                \\ \hline
\end{tabular}
\caption{Calibration data for Fig.~3(b) ibm-manila}
\label{tab:cal2a}
\end{table}

\begin{table}
\begin{tabular}{|c|c|c|c|c|c|c|c|c|c|}
\hline
Qubit & T1 (us) & T2 (us) & Freq. (GHz) & Anharm. (GHz) & Readout error & ID error & $\sqrt{x}$ (sx) error & Pauli-X error & CNOT error                   \\ \hline
Q0    & 81.79   & 143.66  & 4.833       & -0.34189      & 3.090e-2      & 3.181e-4 & 3.181e-4      & 3.181e-4      & 0\_1:9.503e-3                \\ \hline
Q1    & 66.18   & 61.53   & 4.624       & -0.32823      & 2.290e-2      & 3.156e-4 & 3.156e-4      & 3.156e-4      & 1\_2:6.880e-3; 1\_0:9.503e-3 \\ \hline
Q2    & 89.36   & 91.59   & 4.821       & -0.34107      & 1.240e-2      & 1.931e-4 & 1.931e-4      & 1.931e-4      & 2\_3:7.757e-3; 2\_1:6.880e-3 \\ \hline
Q3    & 39.28   & 68.41   & 4.742       & -0.34013      & 6.200e-3      & 1.852e-4 & 1.852e-4      & 1.852e-4      & 3\_4:5.778e-3; 3\_2:7.757e-3 \\ \hline
Q4    & 138     & 161.41  & 4.816       & -0.34291      & 2.070e-2      & 2.294e-4 & 2.294e-4      & 2.294e-4      & 4\_3:5.778e-3                \\ \hline
\end{tabular}
\caption{Calibration data for Fig.~3(c) ibm-santiago}
\label{tab:santiago2b}
\end{table}

\begin{table}
\begin{tabular}{|c|c|c|c|c|c|c|c|c|c|}
\hline
Qubit & T1 (us) & T2 (us) & Freq. (GHz) & Anharm. (GHz) & Readout error & ID error & $\sqrt{x}$ (sx) error & Pauli-X error & CNOT error                   \\ \hline
Q0    & 64.96   & 108.36  & 5           & -0.33689      & 2.780e-2      & 1.969e-4 & 1.969e-4      & 1.969e-4      & 0\_1:9.118e-3                \\ \hline
Q1    & 77.63   & 69.33   & 4.85        & -0.32571      & 2.290e-2      & 2.813e-4 & 2.813e-4      & 2.813e-4      & 1\_2:8.288e-3; 1\_0:9.118e-3 \\ \hline
Q2    & 102.91  & 157.17  & 4.783       & -0.34287      & 3.230e-2      & 1.405e-4 & 1.405e-4      & 1.405e-4      & 2\_3:8.477e-3; 2\_1:8.288e-3 \\ \hline
Q3    & 112.07  & 164.82  & 4.858       & -0.32528      & 1.480e-2      & 6.275e-4 & 6.275e-4      & 6.275e-4      & 3\_4:9.054e-3; 3\_2:8.477e-3 \\ \hline
Q4    & 100.48  & 161.15  & 4.978       & -0.33796      & 1.410e-2      & 1.547e-4 & 1.547e-4      & 1.547e-4      & 4\_3:9.054e-3                \\ \hline
\end{tabular}
\caption{Calibration data for Fig.~3(d) ibm-bogota}
\label{tab:bogota2c}
\end{table}

\clearpage
\section{Symmetric Trotter decomposition for a three-spin chain, its convergence, and additional noisy quantum simulator results}
\label{App:3spinsConvg}

\begin{figure}[ht]
	\centering
		\includegraphics[width=15.0cm]{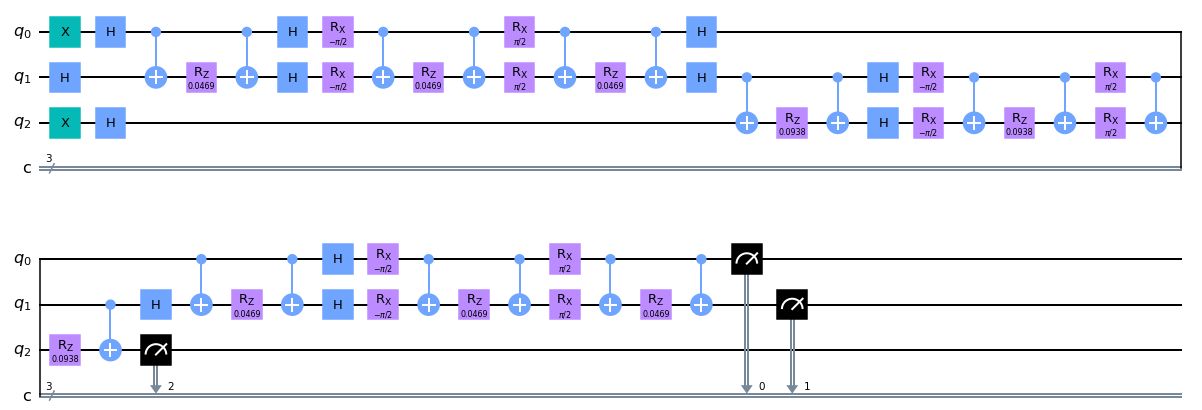}
		\caption{(Color online) User-designed quantum circuit used to simulate a light-driven three-spin chain for time step $n=1$. No Qiskit optimization was implemented at this stage. The parameters are the same as in Fig. 4 in the main text.}
\label{fig:appC1}
\end{figure}

\begin{figure}[ht]
	\centering
		\includegraphics[width=8.5cm]{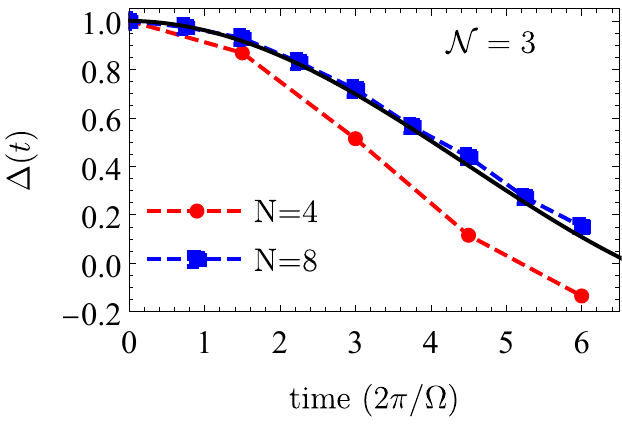}
		\caption{(Color online) Antiferromagnetic order parameter $\Delta(t)$, Eq.\eqref{eq:Delta}, as a function of time, obtained in a clean quantum simulator for two different number of Trotter steps $N=4$ (red circles), and $N=8$ (blue squares) showing the convergence of the symmetric Trotter decomposition. The parameters are the same as in Fig. 4 in the main text.}
\label{fig:app_N_3}
\end{figure}

\begin{figure}[ht]
	\centering
		\includegraphics[width=8.5cm]{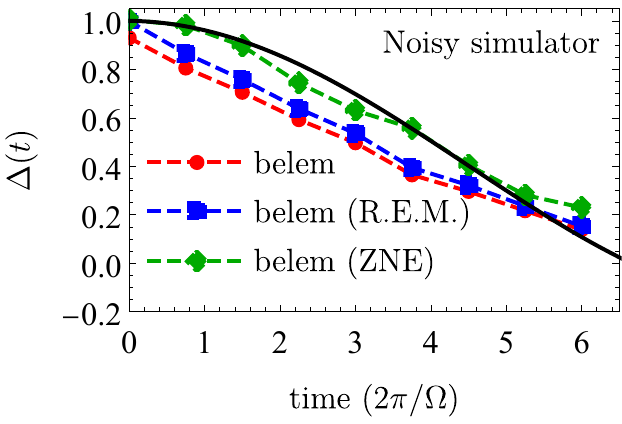}
		\caption{(Color online) AFM order parameter $\Delta(t)$, Eq.\eqref{eq:Delta}, as a function of time for a light-driven three-spin chain with $U=10$, $A=2.8$, and $\Omega=6$. The results were obtained in the ibm-belem noisy simulator. The black line is the effective Floquet exact solution, the red circles are the direct results from the noisy simulator. The blue squares include readout error mitigation (R.E.M), and the green diamonds incorporate zero-noise extrapolation (ZNE). }
\label{fig:spin3noisybelem}
\end{figure}

\begin{table}
\begin{tabular}{|c|c|c|c|c|c|}
\hline
              & $T_1 (\mu s)$ & $T_2 (\mu s)$ & Readout error & X-gate error & CNOT error \\ \hline
Fake Santiago & 124.04        & 107.11        & 0.014         & 0.0002       & 0.00602    \\ \hline
\end{tabular}
\caption{Average calibration data for the noisy simulators used in Fig.~\ref{fig:spin3noisybelem}}
\label{tab:calibration-data-fakedevices3}
\end{table}

\clearpage
\section{Quantum device calibration data for three-spin chains}

\begin{table}[ht]
\begin{tabular}{|c|c|c|c|c|c|c|c|c|c|}
\hline
Qubit & T1 (us) & T2 (us) & Freq. (GHz) & Anharm. (GHz) & Readout error & ID error & $\sqrt{x}$ (sx) error & Pauli-X error & CNOT error                                                                                \\ \hline
Q0    & 82.44   & 141.97  & 5.09        & -0.33612      & 1.800e-2      & 2.025e-4 & 2.025e-4      & 2.025e-4      & 0\_1:2.056e-2                                                                             \\ \hline
Q1    & 109.77  & 121.18  & 5.245       & -0.31657      & 2.700e-2      & 3.053e-4 & 3.053e-4      & 3.053e-4      & \begin{tabular}[c]{@{}c@{}}1\_3:6.573e-3; \\ 1\_2:7.801e-3; \\ 1\_0:2.056e-2\end{tabular} \\ \hline
Q2    & 103.16  & 55.85   & 5.361       & -0.33063      & 1.890e-2      & 2.774e-4 & 2.774e-4      & 2.774e-4      & 2\_1:7.801e-3                                                                             \\ \hline
Q3    & 101.35  & 225.39  & 5.17        & -0.33374      & 1.440e-2      & 2.411e-4 & 2.411e-4      & 2.411e-4      & 3\_4:7.526e-3; 3\_1:6.573e-3                                                              \\ \hline
Q4    & 89.59   & 142.45  & 5.258       & -0.33135      & 1.780e-2      & 1.997e-4 & 1.997e-4      & 1.997e-4      & 4\_3:7.526e-3                                                                             \\ \hline
\end{tabular}
\caption{Calibration data for Fig.~4(c) ibm-belem}
\label{tab:ibm-belem}
\end{table}

\begin{table}[h]
\begin{tabular}{|c|c|c|c|c|c|c|c|c|c|}
\hline
Qubit & T1 (us) & T2 (us) & Freq. (GHz) & Anharm. (GHz) & Readout error & ID error & $\sqrt{x}$ (sx) error & Pauli-X error & CNOT error                                                                                \\ \hline
Q0    & 67.38   & 134.22  & 5.301       & -0.33148      & 3.660e-2      & 2.696e-4 & 2.696e-4      & 2.696e-4      & 0\_1:5.790e-3                                                                             \\ \hline
Q1    & 100.12  & 138.77  & 5.081       & -0.31925      & 1.570e-2      & 2.628e-4 & 2.628e-4      & 2.628e-4      & \begin{tabular}[c]{@{}c@{}}1\_3:9.519e-3; \\ 1\_2:8.301e-3; \\ 1\_0:5.790e-3\end{tabular} \\ \hline
Q2    & 113.12  & 160.77  & 5.322       & -0.33232      & 2.480e-2      & 4.777e-4 & 4.777e-4      & 4.777e-4      & 2\_1:8.301e-3                                                                             \\ \hline
Q3    & 90.46   & 22.67   & 5.164       & -0.33508      & 5.080e-2      & 7.763e-4 & 7.763e-4      & 7.763e-4      & 3\_4:1.665e-2; 3\_1:9.519e-3                                                              \\ \hline
Q4    & 101.25  & 143.36  & 5.052       & -0.31926      & 2.610e-2      & 6.458e-4 & 6.458e-4      & 6.458e-4      & 4\_3:1.665e-2                                                                             \\ \hline
\end{tabular}
\caption{Calibration data for Fig.~4(d) ibm-quito}
\label{tab:tab3d}
\end{table}

\clearpage
\section{Four-spin chain results}

\begin{figure}[ht]
	\centering
		\includegraphics[width=10.5cm]{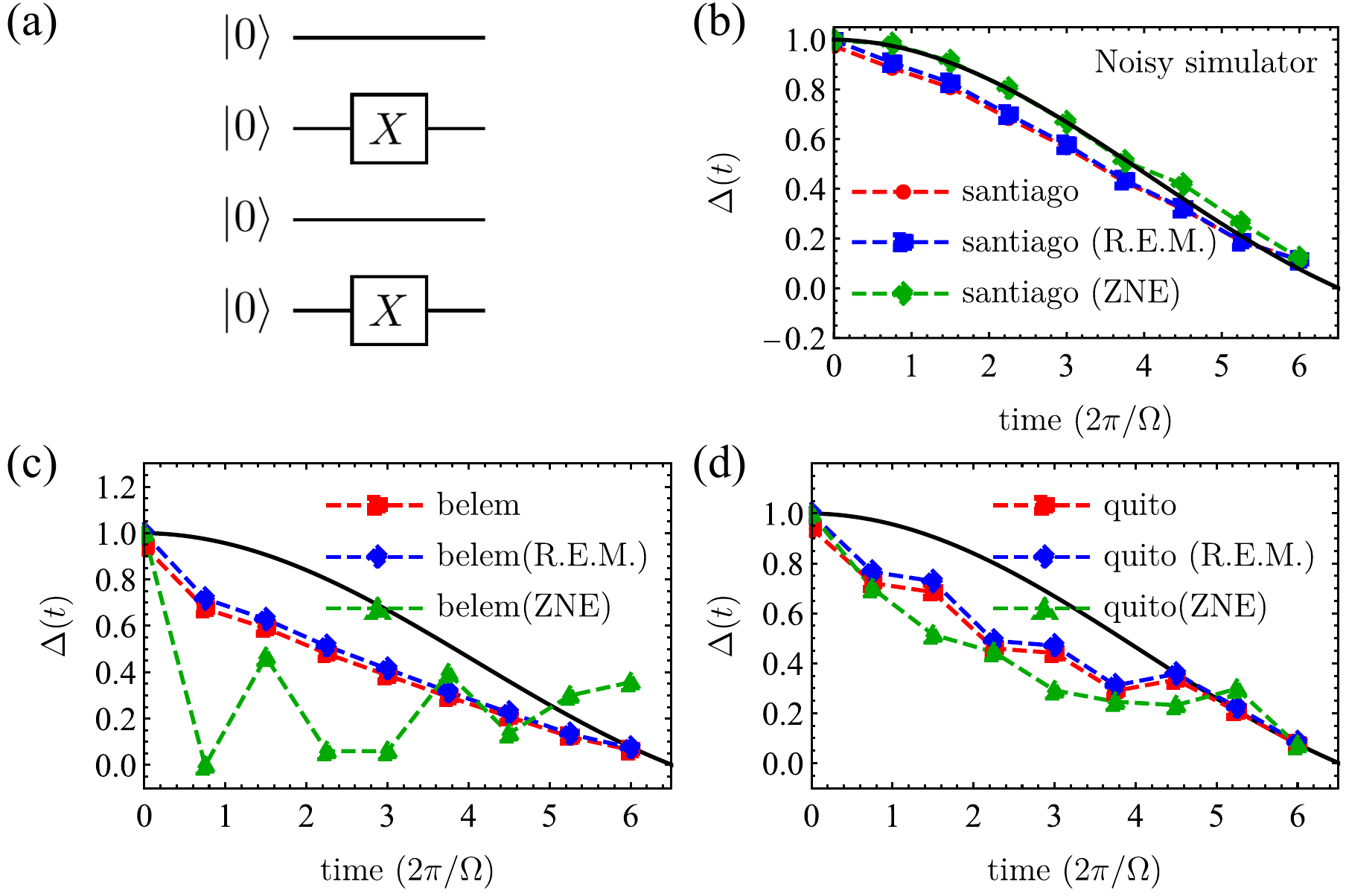}
		\caption{(Color online) (a) Quantum circuit defining the four-site AFM initial state, created with the Quantikz package~\cite{kay_2018}. (b) AFM order parameter $\Delta(t)$, Eq.\eqref{eq:Delta}, as a function of time for a light-driven four-spin chain with $U=10$, $A=2.8$, and $\Omega=6$. The results were obtained in a noisy simulator. The black line is the effective Floquet exact solution, the red circles are the direct results from the noisy simulator. The blue squares include readout error mitigation (R.E.M), and the green diamonds incorporate zero-noise extrapolation (ZNE). Panels (c) and (d) show the corresponding results obtained in the actual quantum devices ibm-belem and ibm-quito.}
\label{fig:figAppFourSpinChain}
\end{figure}

Fig. \ref{fig:figAppFourSpinChain}(a) shows the results for a light-driven spin chains with four spins obtain in an IBM noisy simulator. The bare results (red circles) follow the trend of the exact Floquet solution well, and the zero-noise extrapolation procedure improves the results (green diamonds). The results we obtain in the ibm-belem and ibm-quito devices are shown in Figs.\ref{fig:figAppFourSpinChain}(b) and (c), respectively. The quality of the solutions decrease compared with the three-spin chain counterparts. However, the ibm-quito shows results that approximately follow the trend of the exact solution. We should notice that the zero-noise extrapolation implementation on the quantum devices does not improve the solution as in the three-spin chain case. This could be due to the the large number of CNOT operations required to simulate the additional noise in the extrapolation procedure. The calibration data for the noisy simulator, and quantum devices at the time they were accessed are shown in Tables \ref{tab:calibration-data-fakedevices4}, \ref{tab:belem4spins}, and \ref{tab:quito4spins}.

\begin{table}[ht]
\begin{tabular}{|c|c|c|c|c|c|}
\hline
              & $T_1 (\mu s)$ & $T_2 (\mu s)$ & Readout error & X-gate error & CNOT error \\ \hline
Fake Santiago & 124.04        & 107.11        & 0.014         & 0.0002       & 0.00602    \\ \hline
\end{tabular}
\caption{Average calibration data for the noisy simulators used in Fig.~\ref{fig:figAppFourSpinChain}(b)}
\label{tab:calibration-data-fakedevices4}
\end{table}

\begin{table}[]
\begin{tabular}{|c|c|c|c|c|c|c|c|c|c|}
\hline
Qubit & T1 (us) & T2 (us) & Freq. (GHz) & Anharm. (GHz) & Readout error & ID error & $\sqrt{x}$ (sx) error & Pauli-X error & CNOT error                                                                                \\ \hline
Q0    & 121.16  & 92.24   & 5.09        & -0.33612      & 1.840e-2      & 2.697e-4 & 2.697e-4      & 2.697e-4      & 0\_1:1.418e-2                                                                             \\ \hline
Q1    & 97.63   & 76.2    & 5.246       & -0.31657      & 2.650e-2      & 3.030e-4 & 3.030e-4      & 3.030e-4      & \begin{tabular}[c]{@{}c@{}}1\_3:8.012e-3; \\ 1\_2:6.480e-3; \\ 1\_0:1.418e-2\end{tabular} \\ \hline
Q2    & 107.32  & 50.3    & 5.361       & -0.33063      & 2.310e-2      & 2.363e-4 & 2.363e-4      & 2.363e-4      & 2\_1:6.480e-3                                                                             \\ \hline
Q3    & 127.28  & 160     & 5.17        & -0.33374      & 2.410e-2      & 2.467e-4 & 2.467e-4      & 2.467e-4      & \begin{tabular}[c]{@{}c@{}}3\_4:8.481e-3; \\ 3\_1:8.012e-3\end{tabular}                   \\ \hline
Q4    & 104.92  & 183.28  & 5.258       & -0.33135      & 1.850e-2      & 2.034e-4 & 2.034e-4      & 2.034e-4      & 4\_3:8.481e-3                                                                             \\ \hline
\end{tabular}
\caption{Calibration data for Fig. \ref{fig:figAppFourSpinChain}(c) ibm-belem.}
\label{tab:belem4spins}
\end{table}

\begin{table}[]
\begin{tabular}{|c|c|c|c|c|c|c|c|c|c|}
\hline
Qubit & T1 (us) & T2 (us) & Freq. (GHz) & Anharm. (GHz) & Readout error & ID error & $\sqrt{x}$ (sx) error & Pauli-X error & CNOT error                                                                                \\ \hline
Q0    & 83.39   & 121.06  & 5.301       & -0.33148      & 3.780e-2      & 2.770e-4 & 2.770e-4      & 2.770e-4      & 0\_1:8.412e-3                                                                             \\ \hline
Q1    & 105.55  & 140.44  & 5.081       & -0.31925      & 2.080e-2      & 4.862e-4 & 4.862e-4      & 4.862e-4      & \begin{tabular}[c]{@{}c@{}}1\_3:1.182e-2; \\ 1\_2:1.025e-2; \\ 1\_0:8.412e-3\end{tabular} \\ \hline
Q2    & 96.61   & 125.88  & 5.322       & -0.33232      & 2.110e-2      & 2.598e-4 & 2.598e-4      & 2.598e-4      & 2\_1:1.025e-2                                                                             \\ \hline
Q3    & 157.25  & 21.74   & 5.164       & -0.33508      & 2.650e-2      & 3.022e-4 & 3.022e-4      & 3.022e-4      & \begin{tabular}[c]{@{}c@{}}3\_4:1.537e-2; \\ 3\_1:1.182e-2\end{tabular}                   \\ \hline
Q4    & 60.08   & 80.61   & 5.052       & -0.31926      & 3.230e-2      & 4.543e-4 & 4.543e-4      & 4.543e-4      & 4\_3:1.537e-2                                                                             \\ \hline
\end{tabular}
\caption{Calibration data for Fig. \ref{fig:figAppFourSpinChain}(d) ibm-quito.}
\label{tab:quito4spins}
\end{table}

\clearpage
\section{Five-spin chain results}

\begin{figure}[ht]
	\centering
		\includegraphics[width=8.5cm]{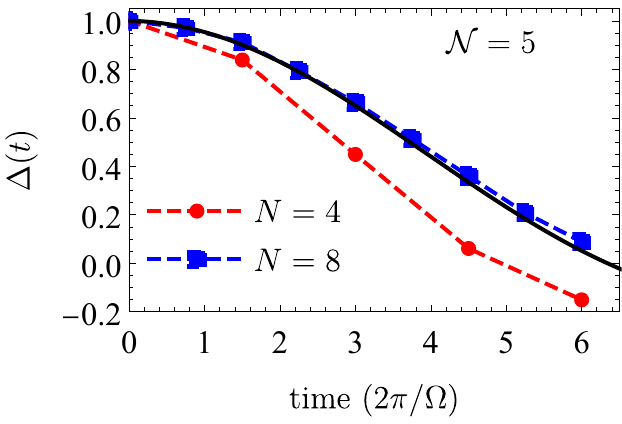}
		\caption{(Color online) Antiferromagnetic order parameter $\Delta(t)$, Eq.\eqref{eq:Delta}, as a function of time, obtained for the two numbers
		of Trotter steps $N=4$ (red circles), and $N=8$ (blue squares) showing the convergence of the symmetric Trotter decomposition in a clean quantum simulator.}
\label{fig:convg_N_5}
\end{figure}

\begin{figure}[ht]
	\centering
		\includegraphics[width=8.5cm]{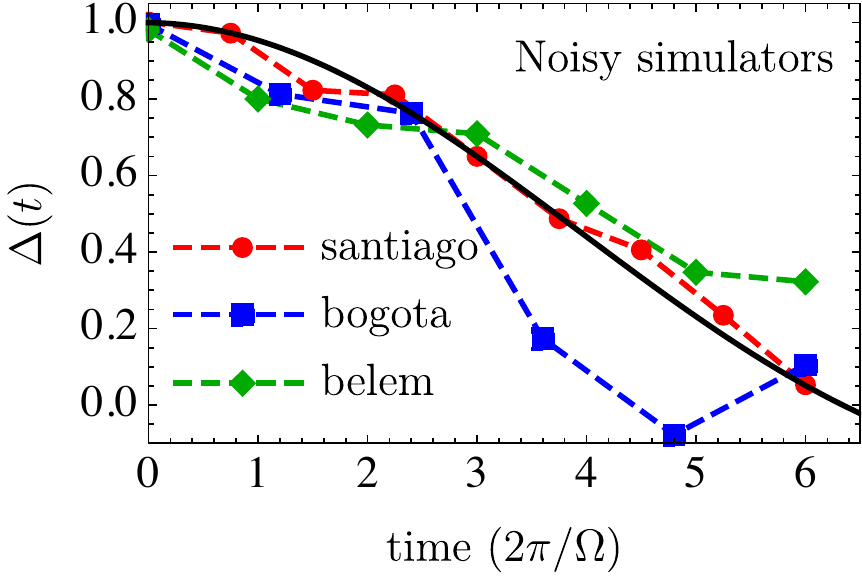}
		\caption{(Color online) Antiferromagnetic order parameter $\Delta(t)$, Eq.\eqref{eq:Delta}, as a function of time, obtained in three different noisy quantum simulators, employing readout error mitigation and zero noise extrapolation in all the cases, as described in the main text. We use $\mathcal{N}=8$ symmetric Trotter steps. The calibration data is shown in Table \ref{tab:calibration-data-fakedevices5}. }
\label{fig:app_N_5}
\end{figure}

In this appendix, we show results for light-driven spin chains with five spins. In Fig. \ref{fig:convg_N_5}, we show the convergence of the symmetric Trotter decomposition. In Fig. \ref{fig:app_N_5} we show the results we obtained in IBM noisy simulators and quantum devices. The solutions including readout error mitigation and zero noise extrapolation follow the trend of the exact results for the effective Floquet exchange interactions obtained via exact diagonalization in the device simulators considered. The averaged errors for the device simulators considered are shown in Table \ref{tab:calibration-data-fakedevices5}. As expected, the device with the smallest CNOT error leads to the most accurate results. 

\begin{figure}[ht]
	\centering
		\includegraphics[width=10.5cm]{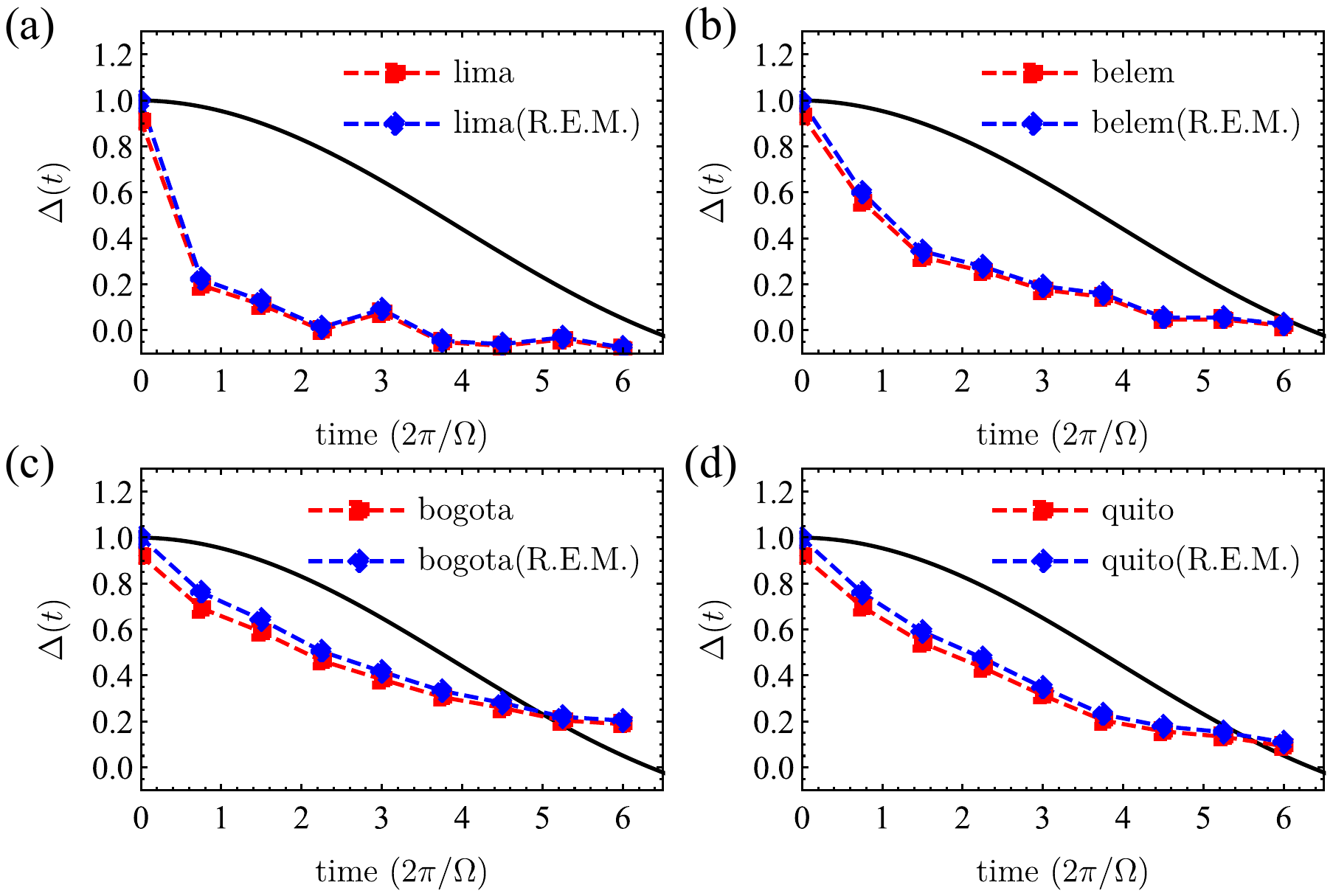}
		\caption{(Color online) (a) AFM order parameter $\Delta(t)$, Eq.\eqref{eq:Delta}, as a function of time for a light-driven five-spin chain with $U=10$, $A=2.8$, and $\Omega=6$. The results were obtained in the ibm-lima quantum device. The black line is the effective Floquet exact solution, the red circles are the direct results from the quantum device. The blue squares include readout error mitigation (R.E.M). Panels (b), (c) and (d) show the corresponding results obtained in the quantum devices ibm-belem, ibm-bogota and ibm-quito.}
\label{fig:figAppFiveSpinChain}
\end{figure}

\begin{figure}[ht]
	\centering
		\includegraphics[width=12.5cm]{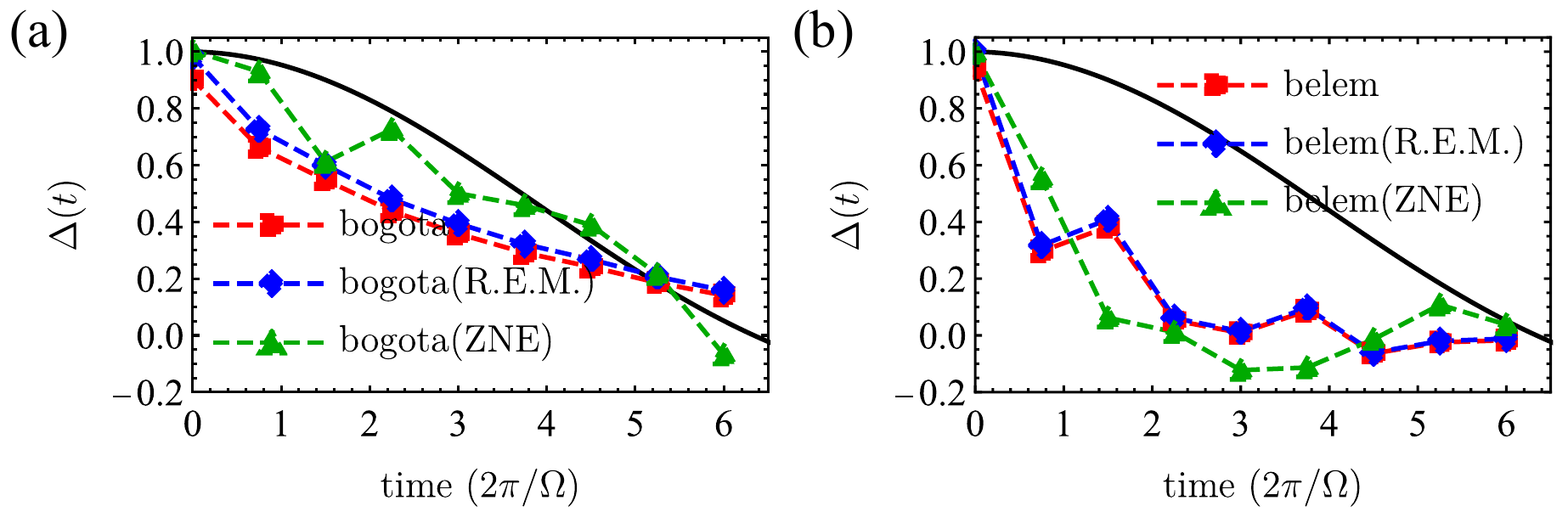}
		\caption{(Color online) AFM order parameter $\Delta(t)$, Eq.\eqref{eq:Delta}, as a function of time for a light-driven five-spin chain with $U=10$, $A=2.8$, and $\Omega=6$. The results were obtained in the (a) ibm-bogota and (b) ibm-belem quantum device. The black line is the effective Floquet exact solution, the red circles are the direct results from the quantum device. The blue squares include readout error mitigation (R.E.M), and the green triangles correspond to results including zero-noise extrapolation. These experiments are independent from Fig. 14 above, obtained in different days.}
\label{fig:figAppFiveSpinChainZNE}
\end{figure}

We should note that once the same quantum circuits are implemented in quantum devices, the quality of the results typically decreases. In Fig. \ref{fig:figAppFiveSpinChain} we show our results from the quantum devices ibm-lima, ibm-belem, ibm-bogota, and ibm-quito. The corresponding calibration data are shown in Tables \ref{tab:fivelima1}, \ref{tab:belemfive1}, \ref{tab:fiveBogota1}, and \ref{tab:quitofive1}. In this case, the results typically do not follow the trend of the exact solution when considering readout error mitigation. However, in some cases, zero-noise extrapolation can improve the quality of the results, as shown in Fig. \ref{fig:figAppFiveSpinChainZNE}. 

\begin{table}[ht]
\begin{tabular}{|c|c|c|c|c|c|}
\hline
              & $T_1 (\mu s)$ & $T_2 (\mu s)$ & Readout error & X-gate error & CNOT error \\ \hline
Fake Santiago & 124.04        & 107.11        & 0.014         & 0.0002       & 0.00602    \\ \hline
Fake Bogota   & 107.56        & 107.025       & 0.0375        & 0.00039      & 0.0284     \\ \hline
Fake Belem  & 80.02        & 79.46         & 0.0304        & 0.00038      & 0.0139    \\ \hline
\end{tabular}
\caption{Average calibration data for the noisy simulators used in Fig.~\ref{fig:app_N_5}}
\label{tab:calibration-data-fakedevices5}
\end{table}

\begin{table}[]
\begin{tabular}{|c|c|c|c|c|c|c|c|c|c|}
\hline
Qubit & T1 (us) & T2 (us) & Freq. (GHz) & Anharm. (GHz) & Readout error & ID error & $\sqrt{x}$ (sx) error & Pauli-X error & CNOT error                                                                                \\ \hline
Q0    & 99.83   & 146.44  & 5.03        & -0.33574      & 1.970e-2      & 1.786e-4 & 1.786e-4      & 1.786e-4      & 0\_1:5.796e-3                                                                             \\ \hline
Q1    & 60.74   & 53.79   & 5.128       & -0.31835      & 1.160e-2      & 5.109e-4 & 5.109e-4      & 5.109e-4      & \begin{tabular}[c]{@{}c@{}}1\_0:5.796e-3; \\ 1\_3:1.568e-2; \\ 1\_2:5.893e-3\end{tabular} \\ \hline
Q2    & 109.16  & 131.69  & 5.247       & -0.3336       & 2.360e-2      & 2.546e-4 & 2.546e-4      & 2.546e-4      & 2\_1:5.893e-3                                                                             \\ \hline
Q3    & 100.13  & 92.88   & 5.302       & -0.33124      & 3.010e-2      & 4.472e-4 & 4.472e-4      & 4.472e-4      & \begin{tabular}[c]{@{}c@{}}3\_4:1.903e-2; \\ 3\_1:1.568e-2\end{tabular}                   \\ \hline
Q4    & 24.62   & 20.65   & 5.092       & -0.33447      & 5.010e-2      & 7.022e-4 & 7.022e-4      & 7.022e-4      & 4\_3:1.903e-2                                                                             \\ \hline
\end{tabular}
\caption{Calibration data for Fig.  14 (a) ibm-lima.}
\label{tab:fivelima1}
\end{table}

\begin{table}[]
\begin{tabular}{|c|c|c|c|c|c|c|c|c|c|}
\hline
Qubit & T1 (us) & T2 (us) & Freq. (GHz) & Anharm. (GHz) & Readout error & ID error & $\sqrt{x}$ (sx) error & Pauli-X error & CNOT error                                                                                \\ \hline
Q0    & 92.62   & 98.19   & 5.09        & -0.33612      & 2.080e-2      & 2.941e-4 & 2.941e-4      & 2.941e-4      & 0\_1:1.518e-2                                                                             \\ \hline
Q1    & 106.84  & 134.98  & 5.246       & -0.31657      & 1.800e-2      & 3.154e-4 & 3.154e-4      & 3.154e-4      & \begin{tabular}[c]{@{}c@{}}1\_3:8.362e-3; \\ 1\_2:6.504e-3; \\ 1\_0:1.518e-2\end{tabular} \\ \hline
Q2    & 85.63   & 45.61   & 5.362       & -0.33063      & 2.170e-2      & 2.685e-4 & 2.685e-4      & 2.685e-4      & 2\_1:6.504e-3                                                                             \\ \hline
Q3    & 96.04   & 120.96  & 5.17        & -0.33374      & 2.160e-2      & 2.616e-4 & 2.616e-4      & 2.616e-4      & \begin{tabular}[c]{@{}c@{}}3\_4:8.295e-3; \\ 3\_1:8.362e-3\end{tabular}                   \\ \hline
Q4    & 122.87  & 192.35  & 5.258       & -0.33135      & 2.400e-2      & 2.544e-4 & 2.544e-4      & 2.544e-4      & 4\_3:8.295e-3                                                                             \\ \hline
\end{tabular}
\caption{Calibration data for Fig.  14 (b) ibm-belem.}
\label{tab:belemfive1}
\end{table}

\begin{table}[]
\begin{tabular}{|c|c|c|c|c|c|c|c|c|c|}
\hline
Qubit & T1 (us) & T2 (us) & Freq. (GHz) & Anharm. (GHz) & Readout error & ID error & $\sqrt{x}$ (sx) error & Pauli-X error & CNOT error                                                              \\ \hline
Q0    & 4.72    & 8.08    & 5           & -0.33689      & 4.200e-2      & 2.425e-3 & 2.425e-3      & 2.425e-3      & 0\_1:7.008e-2                                                           \\ \hline
Q1    & 93.45   & 46.16   & 4.85        & -0.32571      & 4.910e-2      & 2.065e-4 & 2.065e-4      & 2.065e-4      & \begin{tabular}[c]{@{}c@{}}1\_2:8.387e-3; \\ 1\_0:7.008e-2\end{tabular} \\ \hline
Q2    & 110.39  & 176.92  & 4.783       & -0.34287      & 2.910e-2      & 1.299e-4 & 1.299e-4      & 1.299e-4      & \begin{tabular}[c]{@{}c@{}}2\_3:3.339e-2; \\ 2\_1:8.387e-3\end{tabular} \\ \hline
Q3    & 113.4   & 182.15  & 4.858       & -0.32528      & 3.220e-2      & 1.391e-3 & 1.391e-3      & 1.391e-3      & \begin{tabular}[c]{@{}c@{}}3\_4:1.871e-2; \\ 3\_2:3.339e-2\end{tabular} \\ \hline
Q4    & 95.29   & 104.64  & 4.978       & -0.33796      & 1.800e-2      & 1.819e-4 & 1.819e-4      & 1.819e-4      & 4\_3:1.871e-2                                                           \\ \hline
\end{tabular}
\caption{Calibration data for Fig.  14 (c) ibm-bogota.}
\label{tab:fiveBogota1}
\end{table}

\begin{table}[]
\begin{tabular}{|c|c|c|c|c|c|c|c|c|c|}
\hline
Qubit & T1 (us) & T2 (us) & Freq. (GHz) & Anharm. (GHz) & Readout error & ID error & $\sqrt{x}$ (sx) error & Pauli-X error & CNOT error                                                                                \\ \hline
Q0    & 79.46   & 93.63   & 5.301       & -0.33148      & 5.270e-2      & 3.088e-4 & 3.088e-4      & 3.088e-4      & 0\_1:8.347e-3                                                                             \\ \hline
Q1    & 99.77   & 104.96  & 5.081       & -0.31925      & 1.720e-2      & 4.558e-4 & 4.558e-4      & 4.558e-4      & \begin{tabular}[c]{@{}c@{}}1\_3:9.875e-3; \\ 1\_2:8.696e-3; \\ 1\_0:8.347e-3\end{tabular} \\ \hline
Q2    & 85.39   & 136.81  & 5.322       & -0.33232      & 2.170e-2      & 2.292e-4 & 2.292e-4      & 2.292e-4      & 2\_1:8.696e-3                                                                             \\ \hline
Q3    & 112.37  & 21.74   & 5.164       & -0.33508      & 2.120e-2      & 2.532e-4 & 2.532e-4      & 2.532e-4      & \begin{tabular}[c]{@{}c@{}}3\_4:1.434e-2; \\ 3\_1:9.875e-3\end{tabular}                   \\ \hline
Q4    & 52.9    & 89.24   & 5.052       & -0.31926      & 2.850e-2      & 5.019e-4 & 5.019e-4      & 5.019e-4      & 4\_3:1.434e-2                                                                             \\ \hline
\end{tabular}
\caption{Calibration data for Fig.  14 (d) ibm-quito.}
\label{tab:quitofive1}
\end{table}

\begin{table}[]
\begin{tabular}{|c|c|c|c|c|c|c|c|c|c|}
\hline
Qubit & T1 (us) & T2 (us) & Freq. (GHz) & Anharm. (GHz) & Readout error & ID error & $\sqrt{x}$ (sx) error & Pauli-X error & CNOT error                                                              \\ \hline
Q0    & 3.45    & 7.76    & 5           & -0.33689      & 3.050e-2      & 2.370e-3 & 2.370e-3      & 2.370e-3      & 0\_1:5.180e-2                                                           \\ \hline
Q1    & 103.57  & 42.41   & 4.85        & -0.32571      & 5.460e-2      & 5.549e-4 & 5.549e-4      & 5.549e-4      & \begin{tabular}[c]{@{}c@{}}1\_2:1.080e-2; \\ 1\_0:5.180e-2\end{tabular} \\ \hline
Q2    & 108.98  & 192.1   & 4.783       & -0.34287      & 2.330e-2      & 1.541e-4 & 1.541e-4      & 1.541e-4      & \begin{tabular}[c]{@{}c@{}}2\_3:2.504e-2; \\ 2\_1:1.080e-2\end{tabular} \\ \hline
Q3    & 110.86  & 96.48   & 4.858       & -0.32528      & 2.890e-2      & 3.493e-4 & 3.493e-4      & 3.493e-4      & \begin{tabular}[c]{@{}c@{}}3\_4:7.986e-3; \\ 3\_2:2.504e-2\end{tabular} \\ \hline
Q4    & 59.28   & 105.97  & 4.978       & -0.33796      & 2.010e-2      & 1.677e-4 & 1.677e-4      & 1.677e-4      & 4\_3:7.986e-3                                                           \\ \hline
\end{tabular}
\caption{Calibration data for Fig.  15 (a) ibm-bogota.}
\label{tab:fivebogota2}
\end{table}

\begin{table}[]
\begin{tabular}{|c|c|c|c|c|c|c|c|c|c|}
\hline
Qubit & T1 (us) & T2 (us) & Freq. (GHz) & Anharm.y (GHz) & Readout error & ID error & $\sqrt{x}$ (sx) error & Pauli-X error & CNOT error                                                                                \\ \hline
Q0    & 115.11  & 103.17  & 5.09        & -0.33612       & 2.010e-2      & 2.208e-4 & 2.208e-4      & 2.208e-4      & 0\_1:1.146e-2                                                                             \\ \hline
Q1    & 100.05  & 108.58  & 5.246       & -0.31657       & 2.860e-2      & 2.857e-4 & 2.857e-4      & 2.857e-4      & \begin{tabular}[c]{@{}c@{}}1\_3:8.459e-3; \\ 1\_2:7.371e-3; \\ 1\_0:1.146e-2\end{tabular} \\ \hline
Q2    & 72.74   & 49.29   & 5.361       & -0.33063       & 2.180e-2      & 2.748e-4 & 2.748e-4      & 2.748e-4      & 2\_1:7.371e-3                                                                             \\ \hline
Q3    & 131.65  & 130.95  & 5.17        & -0.33374       & 1.970e-2      & 2.721e-4 & 2.721e-4      & 2.721e-4      & \begin{tabular}[c]{@{}c@{}}3\_4:8.978e-3; \\ 3\_1:8.459e-3\end{tabular}                   \\ \hline
Q4    & 120.77  & 172.49  & 5.258       & -0.33135       & 2.660e-2      & 2.312e-4 & 2.312e-4      & 2.312e-4      & 4\_3:8.978e-3                                                                             \\ \hline
\end{tabular}
\caption{Calibration data for Fig.  15 (b) ibm-belem.}
\label{tab:belem15}
\end{table}

\clearpage
\section{Quantum device calibration data for phonon-driven 2 spin chains}

\begin{table}[th]
\begin{tabular}{|c|c|c|c|c|c|c|c|c|c|}
\hline
Qubit & T1 (us) & T2 (us) & Freq. (GHz) & Anharm. (GHz) & Readout error & ID error & $\sqrt{x}$ (sx) error & Pauli-X error & CNOT error                                                                                \\ \hline
Q0    & 124.54  & 143.73  & 5.03        & -0.33574      & 2.290e-2      & 1.907e-4 & 1.907e-4      & 1.907e-4      & 0\_1:5.074e-3                                                                             \\ \hline
Q1    & 109.3   & 99.63   & 5.128       & -0.31835      & 1.540e-2      & 2.539e-4 & 2.539e-4      & 2.539e-4      & \begin{tabular}[c]{@{}c@{}}1\_0:5.074e-3; \\ 1\_3:1.253e-2; \\ 1\_2:6.728e-3\end{tabular} \\ \hline
Q2    & 77.29   & 144.87  & 5.247       & -0.3336       & 4.330e-2      & 5.272e-4 & 5.272e-4      & 5.272e-4      & 2\_1:6.728e-3                                                                             \\ \hline
Q3    & 99.16   & 98.2    & 5.303       & -0.33124      & 2.610e-2      & 2.693e-4 & 2.693e-4      & 2.693e-4      & 3\_4:1.647e-2; 3\_1:1.253e-2                                                              \\ \hline
Q4    & 23.81   & 22.12   & 5.092       & -0.33447      & 4.580e-2      & 6.226e-4 & 6.226e-4      & 6.226e-4      & 4\_3:1.647e-2                                                                             \\ \hline
\end{tabular}
\caption{Calibration data for Fig.~5 ibm-lima}
\label{tab:lima}
\end{table}

\end{document}